\documentclass[aps,prl,amsmath,amssymb,reprint,superscriptaddress,preprintnumbers,showpacs,intlimits]{revtex4-1}
\usepackage{bm,latexsym,mathrsfs,enumerate,color}
\usepackage[mathcal]{euscript}
\usepackage[breaklinks=true,unicode=true,urlcolor = blue,colorlinks = true,citecolor = blue,linkcolor = blue]{hyperref}
\usepackage{graphicx}
\usepackage{pgf}
\usepackage[textsize=scriptsize,shadow,textwidth=16mm]{todonotes}
\usepackage{mathtools}
\renewcommand{\vec}[1]{\bm{#1}}
\usepackage{soul}
\usepackage[normalem]{ulem}
\begin{document}

\title{Magnetochiral symmetry breaking in a M\"obius ring}

\author{Oleksandr V.~Pylypovskyi}
\affiliation{Taras Shevchenko National University of Kiev, 01601 Kiev, Ukraine}

\author{Volodymyr P.~Kravchuk}
 \email{vkravchuk@bitp.kiev.ua}
 \affiliation{Bogolyubov Institute for Theoretical Physics, 03143 Kiev, Ukraine}

\author{Denis~D.~Sheka}
\affiliation{Taras Shevchenko National University of Kiev, 01601 Kiev, Ukraine}

\author{Denys Makarov}
 \affiliation{Institute for Integrative Nanosciences, IFW Dresden, 01069 Dresden, Germany}

\author{Oliver G.~Schmidt}
 \affiliation{Institute for Integrative Nanosciences, IFW Dresden, 01069 Dresden, Germany}

\author{Yuri Gaididei}
 \affiliation{Bogolyubov Institute for Theoretical Physics, 03143 Kiev, Ukraine}

\date{\today}

%
%

\begin{abstract}
We show that the interaction of the magnetic subsystem of a curved magnet with the magnet curvature results in coupling of a topologically nontrivial magnetization pattern and topology of the object. The mechanism of this coupling is explored and illustrated by an example of ferromagnetic M{\"o}bius ring, where a topologically induced domain wall appears as a ground state in case of strong easy-normal anisotropy. For the M\"obius geometry the curvilinear form of the exchange interaction produces an additional effective Dzyaloshinskii-like term which leads to the coupling of the magnetochirality of the domain wall and chirality of the M\"obius ring. Two types of domain walls are found, transversal and longitudinal, which are oriented across and along the M\"obius ring, respectively. In both cases the effect of magnetochirality symmetry breaking is established. The dependence of the ground state of the M\"obius ring on its geometrical parameters and on the value of the easy-normal anisotropy is explored numerically.
\end{abstract}


\maketitle

Curvature driven modification of  physical properties of systems with nontrivial geometry is the subject of intensive research in various areas of physics \cite{Bowick09,Turner10}, e.g. electronic properties of graphene \cite{Juan11}, molecular alignment in liquid crystals \cite{Napoli12}, physics of superconductors \cite{Vitelli04,Fomin12}, macromolecular structures \cite{Forgan11}. In physics of nanomagnetism it is possible to distinguish two groups of the curvature induced effects, namely, magnetochiral effects \cite{Dietrich08,Yan12,Kravchuk12a} and topologically induced magnetization patterning \cite{Yoneya08,Kravchuk12a}. The first family units the phenomena of curvature induced chiral symmetry breaking \cite{Hertel13a}. The latter typically originates from the Dzyaloshinskii-like term appearing in curvilinear form of the exchange interaction \cite{Gaididei14}. Effects of the second group appear in curvilinear magnets, where orientation the anisotropy axis is determined by the geometry, e.g. along normal direction.  Thereby magnetic vortices appear in ground state of spherical magnetic shell with easy-surface anisotropy \cite{Kravchuk12a}, domain walls appear in ground state of a M\"obius ring with easy-normal anisotropy \cite{Yoneya08}.

Up to now these two families of effects were considered to be independent. Here, we demonstrate that geometry of the M\"obius ring unites these two families of curvature effects; namely the topologically induced magnetization patterns experience the chirality symmetry breaking. The M\"obius ring is a nonorientable surface, hence its topology forces a discontinuity of any normal vector field. This is the origin of the domain structure formation for the M\"obius ring with strong easy-normal anisotropy, similar to the nucleation of disclination lines in chiral nematics \cite{Machon13}. Domain wall for the M\"obius geometry was obtained numerically \cite{Yoneya08} within the model of classical Heisenberg ferromagnet. Constrained by the topology, we name such magnetization structure as topologically induced domain wall. We demonstrate analytically that topological properties of a domain wall on a M\"obius ring depend on topological properties of the underlying surface, namely, the magnetochirality of the domain wall is determined by the chirality of the M\"obius ring. Additionally, using the full scale micromagnetic simulations with magnetostatic interaction taken into account, we confirm these results and also build a ground states diagram for M\"obius rings.

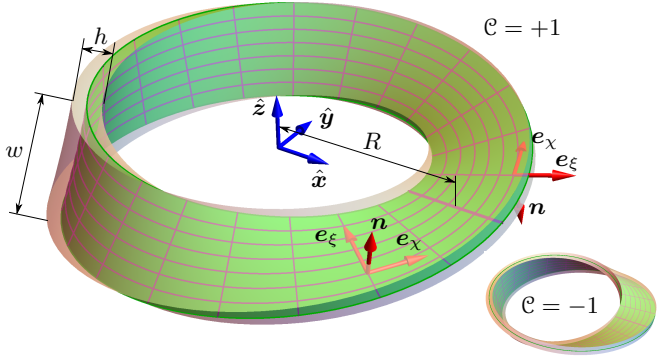
\begin{figure}
\tikz {
\node at (0,0) {\includegraphics[width=\columnwidth]{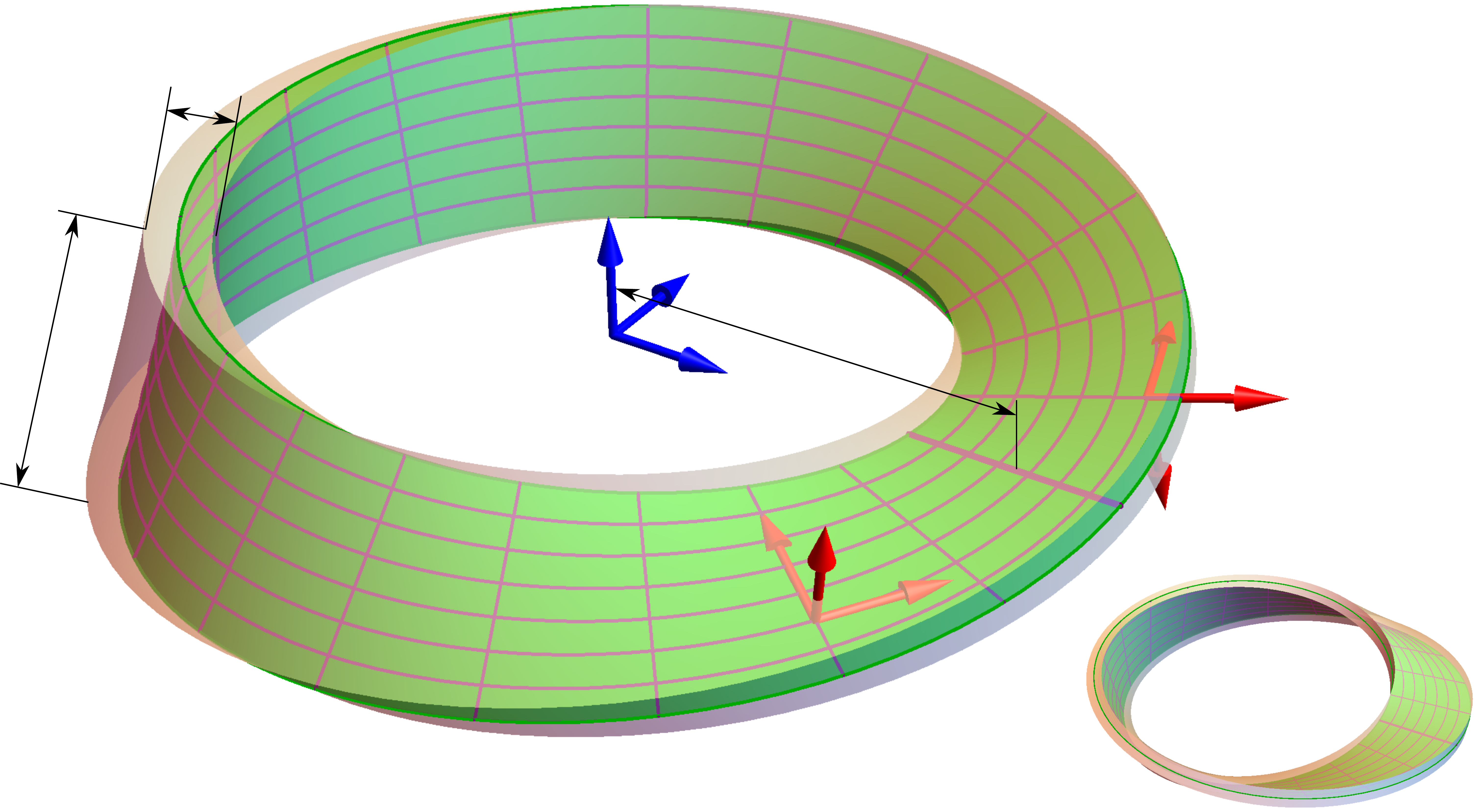}};
\node at (3,-1.7) {$\mathcal{C}=-1$};
\node at (2.5,2) {$\mathcal{C}=+1$};
\node at (-0.1,-0.8) {$\vec e_\xi$};
\node at (1,-0.85) {$\vec e_\chi$};
\node at (0.6,-0.6) {$\vec n$};
\node at (3.1,0.2) {$\vec e_\xi$};
\node at (2.8,0.5) {$\vec e_\chi$};
\node at (2.7,-0.4) {$\vec n$};
\node at (-0.2,0) {$\hat{\vec x}$};
\node at (-0.1,0.8) {$\hat{\vec y}$};
\node at (-1,0.95) {$\hat{\vec z}$};
\node at (0.5,0.5) {$R$};
\node at (-3.1,1.9) {$h$};
\node at (-4.25,0.35) {$w$};
;}
\caption{Geometrical notations of the problem: the 2D M\"obius ring with radius $R$ and width $w$ is defined parametrically by \eqref{eq:parametr2D} and the corresponding 3D M\"obius ring of finite thickness $h$ is determined by \eqref{eq:parameter3D}.  M\"obius rings with opposite chiralities $\mathcal{C}$ are shown.}\label{fig:mobius_notations}
\end{figure}

First, we formalize the notion of the M\"obius ring of finite thickness. The two dimensional M\"obius ring can be parameterized in the form $\vec\varsigma (\chi ,\xi )=x(\chi ,\xi )\hat{\vec x}+y(\chi ,\xi )\hat{\vec y}+z(\chi ,\xi )\hat{\vec z}$ with
\begin{equation} \label{eq:parametr2D}
\begin{aligned}
  &x+iy=\left(R+\xi \cos\frac{\chi }{2}\right)e^{i\chi },&\quad0\le\chi<2\pi,\\
  &z=\mathcal{C}\,\xi \sin\frac{\chi }{2},&\quad-\frac{w}{2}\le\xi \le\frac{w}{2},
\end{aligned}
\end{equation}
where $R$ and $w$ denote radius and width of the ring, respectively, see Fig.~\ref{fig:mobius_notations}. The curvilinear coordinates $\chi$ and $\xi$ correspond to azimuthal angle and position along the ring width, respectively. The M\"obius ring chirality $\mathcal{C}=\pm1$ determines orientation of the ring twist, when moving along the azimuthal direction, namely, counterclockwise ($\mathcal{C}=+1$) or clockwise ($\mathcal{C}=-1$). 

Using \eqref{eq:parametr2D} we introduce the orthonormal curvilinear basis as $\vec e_\alpha=\vec g_\alpha/|\vec g_\alpha|$, where $\vec g_\alpha =\partial_\alpha \vec\varsigma $ and $\alpha \in\{\chi,\xi \}$. The vector of the normal is $\vec n=\vec e_\chi \times\vec e_\xi$, see Fig.~\ref{fig:mobius_notations}.

As a result the three dimensional M\"obius ring of finite thickness $h$ is now defined as the following space domain
\begin{equation} \label{eq:parameter3D}
  \vec r(\chi,\xi,\eta )=\vec\varsigma (\chi,\xi )+\eta \vec n,\qquad -\frac{h}{2}\le\eta \le\frac{h}{2},
\end{equation}
where the third curvilinear coordinate $\eta$ determines the position along the ring thickness. The definition \eqref{eq:parameter3D} is valid for the case of small thickness: $h\ll R-\frac w2$.

To describe the space distribution of the unit magnetization vector $\vec m(\vec r)$, it is convenient to introduce the angular parametrization associated with the local curvilinear basis
\begin{equation} \label{eq:m-loc}
  \vec{m} = \sin\theta \cos\phi \, {\vec{e}}_\chi + \sin\theta \sin\phi \,{\vec{e}}_\xi + \cos\theta \, {\vec{n}},
\end{equation}
with the magnetization angles $\theta =\theta (\chi ,\xi )$ and $\phi =\phi (\chi ,\xi )$.

\begin{figure*}
\tikz {
  \node at (0,0) {\includegraphics[width=\textwidth]{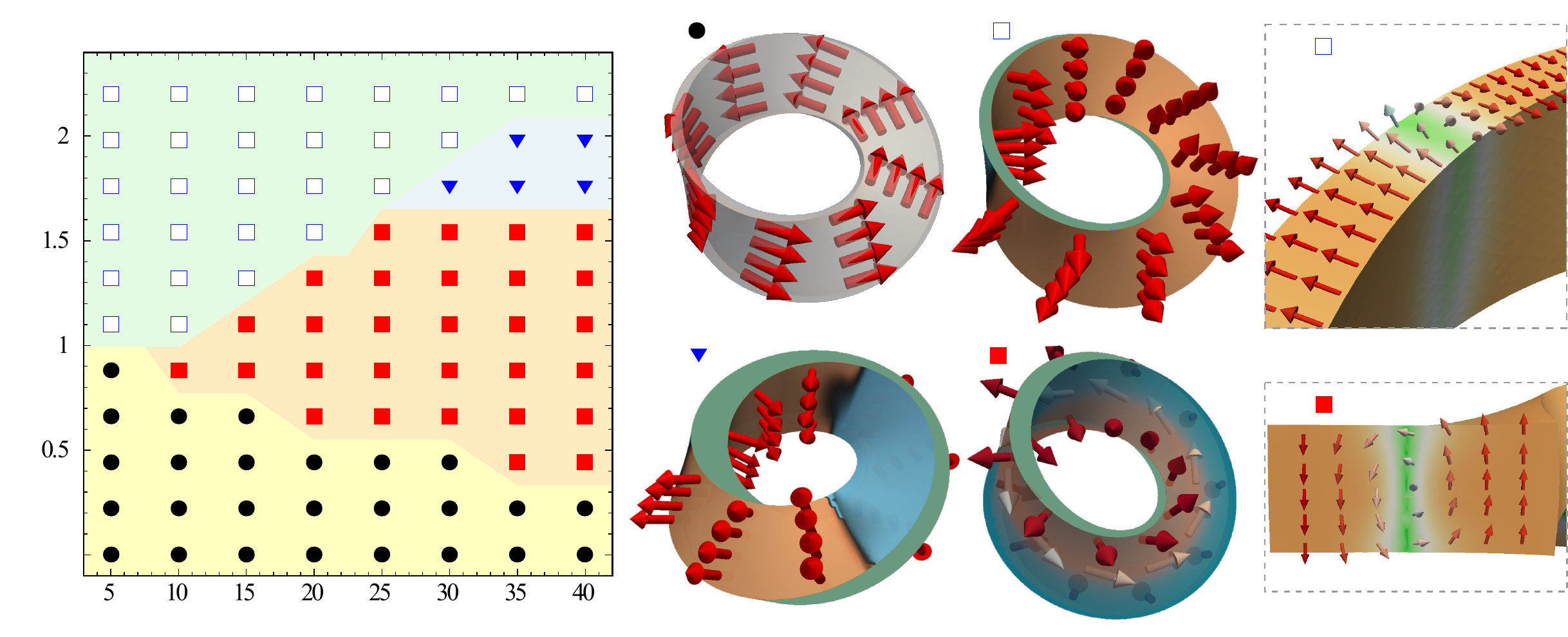}};
  \node at (-5,-3.6) {$h$ (nm)};
  \node[rotate=90] at (-8.7,0) {$Q$};
  \node at (-8.7,3.3) {(a)};
  \node at (-1.35,3.3) {(b)};
  \node at (2.15,3.3) {(c)};
  \node at (-1.35,-0.4) {(d)};
  \node at (2.1,-0.4) {(e)};
  \node at (5.8,3.1) {(c$'$)};
  \node at (5.8,-0.98) {(e$'$)};
;}
\caption{Diagram of ground states of magnetic M\"obius rings with fixed radius $R=100$ nm and width $w=80$, see inset (a). The magnetization distributions of possible ground states are shown in the middle part in a large scale: (b) -- vortex state (marked by disks {\Large $\bullet$} on the diagram), (c) state with single transversal Bloch domain wall (open squares $\textcolor{blue}{\square}$), (d) state with three transversal Bloch walls (filled triangles $\textcolor{blue}{\blacktriangledown}$), (e) states with longitudinal domain wall (filled squares $\textcolor{red}{\blacksquare}$). The detailed structures of the transverse and longitudinal domain walls are shown in the insets (c$'$) and (e$'$) respectively. }\label{fig:mobius_diagram}
\end{figure*}
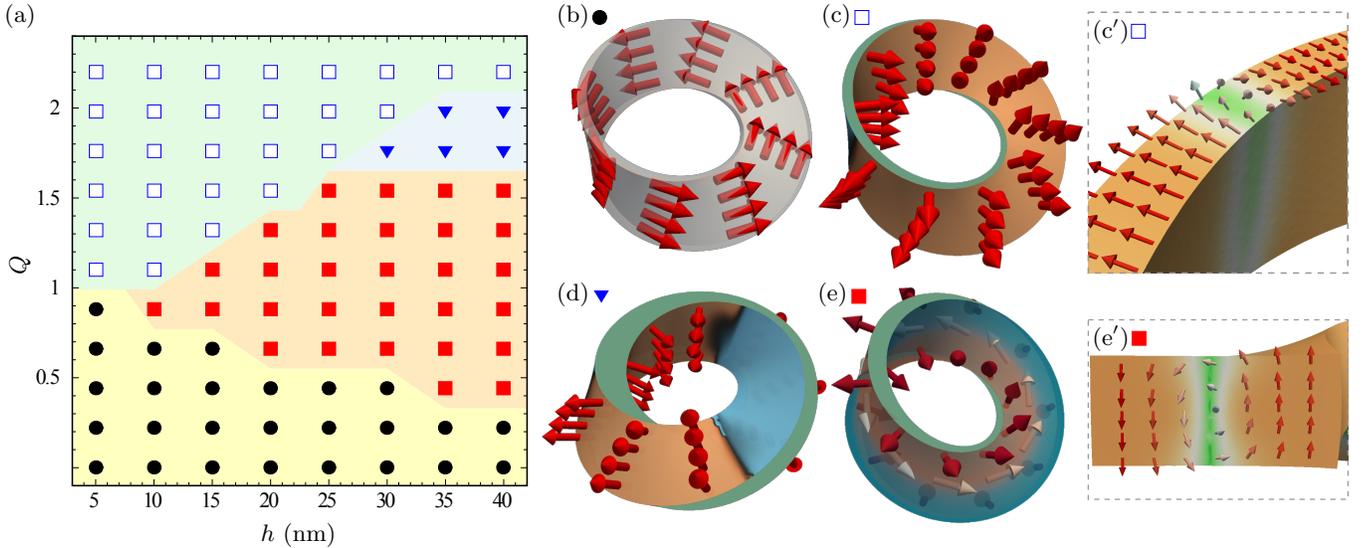

We start with the case of strong easy-normal anisotropy, when the formation of the topologically induced domain walls is expected \cite{Yoneya08}. Two types of domain walls are found using full scale micromagnetic simulations, namely, transversal (\textit{t}-wall) and longitudinal ($\ell$-wall), see Fig.~\ref{fig:mobius_diagram}(c) and Fig.~\ref{fig:mobius_diagram}(e), respectively. We analyse the properties of the topologically induced domain walls under two assumptions: (i) the magnetostatic contribution is negligibly small as compared with the anisotropy contribution, (ii) the ring thickness $h$ is small enough to ensure the magnetization uniformity along the normal direction $\vec n$.  Thus, the total energy of the system reads $E=h\int(\mathcal{E}_{\mathrm{ex}}+\mathcal{E}_{\mathrm{an}})\mathrm{d}\mathcal{S}$, where $\mathrm{d}\mathcal{S}$ is the curvilinear element of the surface area.
Here, $\mathcal{E}_\mathrm{ex}=A\sum_{i=x,y,z} \left(\vec\nabla m_i\right)^2$ is the exchange contribution with $A$ being the exchange constant and $\mathcal{E}_\mathrm{an}=-K(\vec m\cdot\vec n)^2$ is the easy-normal anisotropy density with $K>0$ being the anisotropy constant. In terms of the angular variables \eqref{eq:m-loc} the anisotropy term reads as
\begin{equation} \label{eq:Ean}
\mathcal{E}_\mathrm{an}=-K\cos^2\theta.
\end{equation}
The exchange energy density for an arbitrary curvilinear thin ferromagnetic film can be presented in the form \cite{Gaididei14}
\begin{equation} \label{eq:Eex}
\!\!\! \frac{\mathcal{E}_\mathrm{ex}}{A}=\left[\vec{\nabla}\theta -\vec{\varGamma}(\phi )\right]^2 \!\! + \!\! \left[\sin\theta \left(\vec{\nabla}\!\phi-\vec{\varOmega}\right)\! - \!\cos \theta \frac{\partial \vec{\varGamma}(\phi )}{\partial\phi }\right]^2\!\!\!\!\!.
\end{equation}
Here, the del operator is used in its curvilinear form $\vec\nabla\equiv(g_{\alpha \alpha })^{-1/2}{\vec e}_\alpha \partial_\alpha$, where $g_{\alpha \beta } = \vec{g}_\alpha \cdot \vec{g}_\beta$ are elements of the metric tensor and the Einstein summation rule is applied here and everywhere below. Vector $\vec\varGamma=\varGamma_\alpha\vec e_\alpha$ is tangential to the surface and it is defined as
\begin{equation*} 
  \vec\varGamma(\phi )=\begin{Vmatrix}H_{\alpha \beta }\end{Vmatrix}\cdot\begin{Vmatrix}\cos\phi \\ \sin\phi \end{Vmatrix},\quad \begin{Vmatrix}H_{\alpha \beta }\end{Vmatrix}=\begin{Vmatrix}\frac{b_{\alpha \beta }}{\sqrt{g_{\alpha \alpha }g_{\beta \beta }}}\end{Vmatrix},
\end{equation*}
where $b_{\alpha \beta } = {\vec{n}}\cdot \partial_\beta \vec{g}_\alpha$ are elements of the second fundamental form and the vector $\vec\varOmega$ represents the modified spin connection $\vec\varOmega=(g_{\alpha \alpha })^{-1/2}(\vec{e}_{\chi }\cdot\partial_\alpha \vec{e}_{\xi })\vec{e}_{\alpha }$. 

Basic properties of the \textit{t}-wall can be analysed using the Ansatz
\begin{equation} \label{eq:Ansatz}
  \theta^t=2\arctan e^{p\frac{\chi-X}{\sigma }},\quad\phi^t=\mathfrak{C}p\frac{\pi }{2},
\end{equation}
which describes the structure of a typical Bloch domain wall \cite{Landau35} aligned across the M\"obius ring. It contains two variational parameters, namely, azimuthal angle $X$ which determines the position of the \textit{t}-wall and its angular width $\sigma$. The quantity $p=\pm1$ determines the wall of kink ($p=+1$) or anti-kink ($p=-1$) type. The magnetochirality $\mathfrak{C}=\pm1$ determines the direction of the magnetization reorientation within the domain wall, when moving along the azimuthal direction: counterclockwise ($\mathfrak{C}=+1$) or clockwise ($\mathfrak{C}=-1$). The usage of the Ansatz \eqref{eq:Ansatz} for a M\"obius ring has a restriction $\sigma\ll X\ll2\pi-\sigma$.

Now we substitute Ansatz \eqref{eq:Ansatz} into \eqref{eq:Ean} and \eqref{eq:Eex} and perform the integration over the curvilinear surface area with the area element $\mathrm{d}\mathcal{S}=\sqrt{\det(g_{\alpha \beta })}\mathrm{d}\chi \mathrm{d}\xi$. Using the natural condition $\sigma\ll1$ and applying the narrow ring approximation $w/R\ll1$, one can write the total energy of the \textit{t}-wall as follows
\begin{equation} \label{eq:E-narrow}
\begin{split}
  \frac{E^t}{2Ah}\approx&\frac{w}{R}\left[\frac{1}{\sigma }+\mathfrak{C}\mathcal{C}\frac{\pi }{2}+\sigma \left(\cos X-\frac14\right)\right]+\sigma \frac{Rw}{\delta^2}\\
  &+\mathrm{const},
\end{split}
\end{equation}
where $\delta=\sqrt{A/K}$ and the constant terms do not depend on the variational parameters and on the chiralities.

Remarkably, according to \eqref{eq:E-narrow}, a coupling between the chirality of the M\"obius ring $\mathcal{C}$ and the magnetochirality of the \textit{t}-wall $\mathfrak{C}$ takes place. The \textit{t}-walls with opposite chiralities possess different energies with $\Delta E^\textit{t}=2\pi A hw/R$ (the \textit{t}-wall with chirality $\mathfrak{C}=-\mathcal{C}$ has lower energy). We note, that the energy \eqref{eq:E-narrow} is independent of $p$.

\begin{figure}
\tikz {\node at (0,0) {\includegraphics[width=\columnwidth]{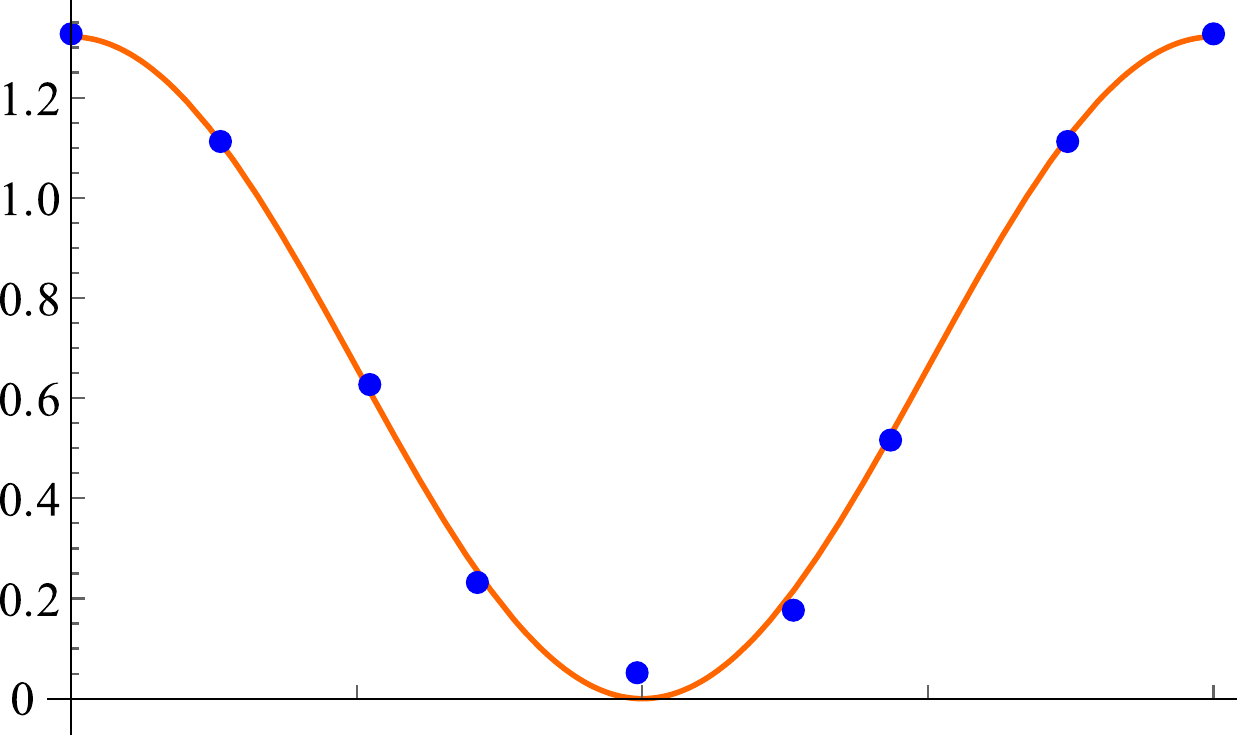}};
  \node at (-3.25,2.8) {$\Delta E/E_0$ ($10^{-3}$)};
  \node at (4.2,-2) {$X$};
  \node at (4.15,-2.55) {$2\pi$};
  \node at (2.2,-2.55) {$3\pi/2$};
  \node at (0.19,-2.55) {$\pi$};
  \node at (-1.8,-2.55) {$\pi/2$};
  \node at (0.19,-1.85) {$X_0$};
;}
\caption{Increase of the total energy $\Delta E$ when the \textit{t}-wall is shifted from its equilibrium position $X_0=\pi$. Points represent the simulation results for the M\"obius ring with $h=5$ nm and $Q=1.72$, solid line shows fitting by the expected dependence $\cos X$, see~\eqref{eq:E-narrow}. The normalizing constant is $E_0=4\pi M_s^2V$ with $V$ being the ring volume. The case of the opposite chiralities $\mathcal{C}=+1$ and $\mathfrak{C}=-1$ is considered.}\label{fig:equilibr_position}
\end{figure}
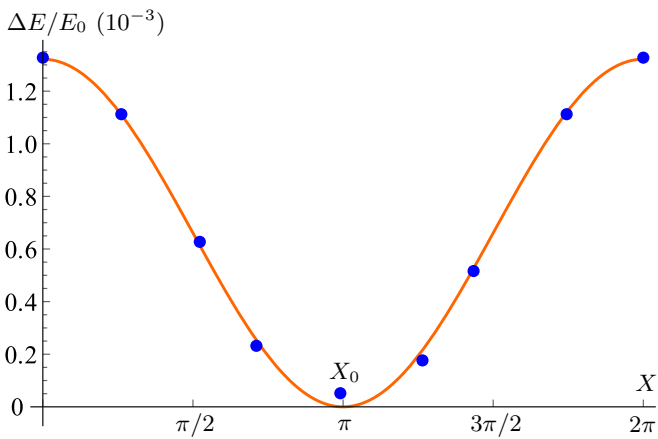

The energy \eqref{eq:E-narrow} reaches its minimum at the following values of the variational parameters $X_0=\pi$, $\sigma_0\approx\delta/R$.
In accordance with the parametrization \eqref{eq:parametr2D}, the equilibrium position of the domain wall $X_0=\pi$ corresponds to the ``vertical'' place on the M\"obius ring, see Fig.~\ref{fig:mobius_notations} and Fig.~\ref{fig:mobius_diagram}(c). This result is confirmed by full scale micromagnetic simulations, see Fig.~\ref{fig:equilibr_position}.


To consider the $\ell$-wall, see Fig~\ref{fig:mobius_diagram}(e), we use the analogous Ansatz
\begin{equation} \label{eq:ansatz-long}
  \theta^\ell=2\arctan e^{p\frac{\xi }{d}},\quad\phi^\ell=\pi \frac{\mathfrak{C}+p}{2},
\end{equation}
which describes a Bloch domain wall aligned along the M\"obius ring. It should be noted that the $\ell$-wall cannot be of Neel type due to topological reasons. The domain wall width $d$ is a variational parameter. Using the condition $d\ll w\ll R$ we obtain the following expression for the total energy of the $\ell$-wall state
\begin{equation} \label{eq:en-long}
  \frac{E^\ell}{4\pi A h}\approx\frac{R}{d}-\frac{\pi }{2}\mathcal{C}\mathfrak{C}+c_0\frac{d}{R}+\frac{d R}{\delta^2}+\mathrm{const},
\end{equation}
where $c_0=1/4+\pi^2/96\approx0.353$. Accordingly to \eqref{eq:en-long}, the chiralities coupling appears for the $\ell$-wall as well as for the case of the \textit{t}-wall. The $\ell$-walls with opposite chiralities are separated by the energy gap $\Delta E^\ell=4\pi^2A h$ (the domain wall with $\mathfrak{C}=\mathcal{C}$ has lower energy). The equilibrium value of the domain wall width is $d\approx\delta$.

For both domain wall types the effect of the magnetochirality symmetry breaking originates from the effective Dzyaloshinskii-like term $\mathcal{E}_\mathrm{ex}^{D} = -2A\left(\vec\varGamma \cdot \vec\nabla\theta \right)$ in the curvilinear form of the exchange energy \eqref{eq:Eex}. 

The curvature induced magnetochirality effects are already known for domain structure (twisting of domains) \cite{Dietrich08}, spin waves (asymmetry in the spin waves propagation in nanotubes) \cite{Hertel13a}. Its analogy with the Dzyaloshinskii-Moriya interaction was discussed in Ref.~\cite{Hertel13a}. The presented approach highlights  the origin of the these magnetochiral effects.

According to \eqref{eq:E-narrow} and \eqref{eq:en-long}, the total energy of a \textit{t}-wall is always lower than the total energy of a $\ell$-wall. However, the carried out analysis is valid for the case of very strong anisotropy, when the magnetostatic interaction can be neglected. To consider cases when the magnetostatic contribution is comparable or greater then the anisotropy interaction, we perform the full scale micromagnetic simulations with three magnetic interactions taken into account. Thus, the total energy density reads $\mathcal{E}=\mathcal{E}_\mathrm{ex}+\mathcal{E}_\mathrm{an}+\mathcal{E}_\mathrm{d}$. Here, $\mathcal{E}_\mathrm{d}=-M_s(\vec H_d\cdot\vec m)/2$ is the magnetostatic energy density, with $M_s$ being the saturation magnetization and $\vec H_d$ being the stray field. To clarify the role of the magnetostatic contribution, we study the change of the ground state of the system under variation of two parameters, namely, thickness $h$ and a quality factor $Q=K/(2\pi M_s^2)$, the latter relates the anisotropy to the magnetostatic contribution \cite{Hubert98}.

For our study we fix the ring radius to $R=100$~nm and width to $w=80$~nm. The calculations are performed for a magnetic material with $A=1.3\times10^{-11}$~J/m, $M_s=8.6\times10^5$~A/m. Additionally we introduce the easy-normal anisotropy with constant $K$ varying \footnote{The largest value of the anisotropy constant is limited by the chosen average mesh size 3 nm which is determined by the computational possibilities.} in the way that $Q\in[0,\,2.2]$. The thickness is varied \footnote{The lower bound corresponds to a quasi 2D case when the magnetization is uniform along normal, and the upper bound is limited by conditions of usage of the definition \eqref{eq:parameter3D}.} within the range $h\in[5,\,40]$~nm. For a certain set of parameters we use the $\mathrm{MAGPAR}$ code \cite{MAGPAR} to minimize numerically the total magnetic energy of the system $E=\int\!\mathcal{E}\,\mathrm{d}^3\vec r$. To obtain the ground state among the variety of possible metastable states, we applied a large number of various initial states \footnote{We use 12 different initial states, namely, four different states with random magnetization distributions, six states uniformly magnetized along directions $\pm\hat{\vec x}$, $\pm\hat{\vec y}$, $\pm\hat{\vec z}$ respectively, and two vortex states with opposite senses of the circulation. The lowest energy state obtained after applying the energy minimization procedure is considered to be the ground state for the certain set of parameters. } for the minimization procedure in each case. The results are presented in Fig.~\ref{fig:mobius_diagram}.

We conclude from Fig.~\ref{fig:mobius_diagram}(a) that the ground state of the system with low anisotropy is a vortex magnetization distribution, with the magnetizaiton vectors tangentially aligned to the M\"obius surface, see Fig.~\ref{fig:mobius_diagram}(b). This is the typical situation for magnetically soft nanomagnets with symmetric shape, where the vortex state dominates as a result of competition between exchange and magnetostatic interactions only, e.g. in magnetic nanodisks \cite{Cowburn99} and nanorings \cite{Klaui03a,*Kravchuk07}. The vortex states with magnetochiralities of opposite signs are energetically equivalent for the M\"obius ring.

In the opposite case of high easy-normal anisotropy the \textit{t}-wall appears as the ground state, see Fig.~\ref{fig:mobius_diagram}(c),(c$'$). According to Fig.~\ref{fig:mobius_diagram}(c$'$), the \textit{t}-wall magnetochirality $\mathfrak{C}=-1$ is opposite to the M\"obius ring chirality $\mathcal{C}=+1$. This confirms the conclusion about the coupling of chiralities of the object and magnetization pattern, which follows from \eqref{eq:E-narrow}. Moreover, change of the M\"obius ring chirality to the opposite one leads to the corresponding switching of the magnetochirality of the \textit{t}-wall \footnote{We checked it for parameters $h=5$ nm, $Q=1.72$}.

By relaxing the \textit{t}-wall in different positions $\chi=X$ along the M\"obius ring, we confirm the conclusion that $E^\textit{t}\propto\cos X$ which results in the equilibrium position $X_0=\pi$, see Fig.~\ref{fig:equilibr_position}.
It should be noted that the \textit{t}-wall with energetically unfavorable magnetochirality $\mathfrak{C}=\mathcal{C}=+1$ flips its magnetochirality during the simulation process. Therefore the corresponding data for the case $\mathcal{C}\mathfrak{C}=+1$ are not shown. However, such a behavior confirms the above conclusion on coupling of chiralitues of the object and its magnetization pattern.

We found that the number of \textit{t}-walls increases with increasing thickness. For instance, states with three \textit{t}-walls  are found, see Fig.~\ref{fig:mobius_diagram}(d). We find out that only odd number of \textit{t}-walls is possible. The appearance of a multidomain structure is the typical consequence of magnetostatic interaction, because the creation of domains leads to stray field energy minimization.

For the M\"obius rings of small thickness the transition between vortex state and \textit{t}-wall state appears for $Q\approx1$. However, for larger thicknesses there is a range of $Q$, where the state with $\ell$-wall is the ground state, see Fig.~\ref{fig:mobius_diagram}(a),(e). According to Fig.~\ref{fig:mobius_diagram}(e) the $\ell$-wall magnetochirality $\mathfrak{C}=+1$ coincides with the M\"obius ring chirality $\mathcal{C}=+1$. This confirms the conclusion about the object-pattern chirality coupling, which follows from \eqref{eq:en-long}. Similarly to the case with \textit{t}-wall, the alternation of the M\"obius ring chirality to opposite one leads to the corresponding flip of the magnetochirality of the $\ell$-wall~\footnote{We checked it for parameters $h=25$ nm, $Q=1.08$.}. The $\ell$-wall is an asymmetric Bloch domain wall \cite{Hubert69,*Hubert70}, see Fig.~\ref{fig:mobius_diagram}(e$'$). The influence of magnetostatics leads to an asymmetric deformation of the Bloch domain wall such that for thick samples the vortex formation along the wall is obtained \cite{Hubert69,*Hubert70}, see Fig.~\ref{fig:mobius_diagram}(e$'$).

In summary, we demonstrate that the combined curvature induced effect appears for M\"obius shaped magnetic nanoring with easy-normal anisotropy, namely the chirality symmetry breaking takes place for both types of the topologically induced domain walls, transversal and longitudinal. This effect is driven by the effective Dzyaloshinskii-like term, which originates from the curvilinear form of the exchange interaction. It is also shown that the ``vertical'' part of the ring is the equilibrium position of the transverse domain wall.

Experimental verification of the predicted effects is foreseeable as a micrometer sized single crystal M\"{o}bius ring is already realized experimentally \cite{Tanda02}. The experimental realization can be addressed to all-electrical measurements by monitoring quantum spin Hall effect \cite{Beugeling14} or topological Hall effect \cite{Porter14} in the M\"obius ring.

This work completes the broad theoretical studies of various physical phenomena for the M\"obius geometry \cite{Nakamura04,Guo09,Guclu13,Jiang10,Wang10,Gravesen05,Li12a,Fomin12a,Yakubo03,Ballon08} by including magnetic phenomena.

We thank to Prof. Vladimir Fomin (IFW Dresden) for inspiriting discussions. This work is financed in part via the European Research Council within the European Union's Seventh Framework Programme (FP7/2007-2013) / ERC grant agreement no. 306277.

%

\end{document}